\documentclass[twocolumn]{aastex63}

\newcommand{\mstar}{$M_{\ast}$}
\newcommand{\msun}{$M_{\odot}$}
\newcommand{\Lsun}{$L_{\odot}$}

\newcommand{\sfruvc}{SFR$_{\rm UV,corr}$}
\newcommand{\sfruvir}{SFR$_{\rm UV+IR}$}
\newcommand{\Luvc}{$L_{\rm NUV,corr}$}
\newcommand{\dlssfr}{$\Delta$\,log\,sSFR}
\newcommand{\delssfr}{$\Delta$\,log\,sSFR}
\newcommand{\dlssfruv}{$\Delta$\,log\,sSFR$_{\rm UV,corr}$}
\newcommand{\dlssfrIR}{$\Delta$\,log\,sSFR$_{\rm UV+24 {\mu}m}$}

\newcommand{\delsma}{$\Delta$\,log\,$R_{\rm e}$}
\newcommand{\dlRe}{$\Delta$\,log\,$R_{\rm e}$}
\newcommand{\Av}{$A_{\rm V}$}

\newcommand{\Anuv}{$A_{\rm NUV}$}
\newcommand{\siggas}{$\Sigma_{\rm gas}$}
\newcommand{\sigsfr}{$\Sigma_{\rm SFR}$}
\newcommand{\sigstar}{$\Sigma_{\ast}$}
\newcommand{\fgas}{$f_{\rm gas}$}

\newcommand{\re}{$R_{\rm e}$}
\newcommand{\reff}{$R_{\rm e}$}
\newcommand{\redshift}{$z$}

\newcommand{\um}{$\mu$m}
\newcommand{\Mvir}{$M_{\rm vir}$}

\newcommand{\Rvir}{$R_{\rm vir}$}
\newcommand{\HI}{{H{\sc \,i}}}
\newcommand{\HII}{{H{\sc \,ii}}}
\newcommand{\MHmol}{$M_{\rm H2}$}
\newcommand{\sersic}{S\'ersic}
\newcommand{\mdotin}{$\dot{M}_{\rm in}$}

\newcommand{\mdots}{$\dot{M}_{\ast}$}

\begin{document}

\title{The SFR-radius connection: data and implications for wind strength and halo concentration}

\author{Lin Lin}
\affiliation{Shanghai Astronomical Observatory, Chinese Academy of Sciences, Shanghai 200030, China}

\author{S.~M.~Faber}
\affiliation{Department of Astronomy and Astrophysics, University of California, Santa Cruz, CA 95064, USA}

\author{David C.~Koo}
\affiliation{Department of Astronomy and Astrophysics, University of California, Santa Cruz, CA 95064, USA}

\author{Samir Salim}
\affiliation{Department of Astronomy, Indiana University, Bloomington, IN 47404, USA}

\author{Aaron A.~Dutton}
\affiliation{New York University Abu Dhabi, PO Box 129188, Saadiyat Island, Abu Dhabi, United Arab Emirates}

\author{Jerome J.~Fang}
\affiliation{Astronomy Department, Orange Coast College, Costa Mesa, CA, 92626, USA}

\author{Fangzhou Jiang}
\affiliation{Racah Institute of Physics, The Hebrew University, Jerusalem 91904, Israel}

\author{Christoph T.~Lee}
\affiliation{Physics Department, University of California, Santa Cruz, CA 95064, USA}

\author{Aldo Rodr\'{i}guez-Puebla}
\affiliation{Instituto de Astronom\'{i}a, Universidad Nacional Aut\'{o}noma de M\'{e}xico, A. P. 70-264, 04510, M\'{e}xico, D.F., M\'{e}xico}

\author{A.~van der Wel}
\affiliation{Sterrenkundig Observatorium, Universiteit Gent, Krijgslaan 281 S9, B-9000 Gent, Belgium}
\affiliation{Max-Planck-Institut f\"{u}r Astronomie, K\"{o}nigstuhl 17, D-69117, Heidelberg, Germany}

\author{Yicheng Guo}
\affiliation{Department of Physics and Astronomy, University of Missouri, Columbia, MO 65211, USA}

\author{Guillermo Barro}
\affiliation{Department of Physics, University of the Pacific, 3601 Pacific Avenue, Stockton, CA 95211, USA}

\author{Joel R.~Primack}
\affiliation{Physics Department, University of California, Santa Cruz, CA 95064, USA}

\author{Avishai Dekel}
\affiliation{Racah Institute of Physics, The Hebrew University, Jerusalem 91904, Israel}
\affiliation{SCIPP, University of California, Santa Cruz, CA 95064, USA}

\author{Zhu Chen}
\affiliation{Shanghai Key Lab for Astrophysics, Shanghai Normal University, 100 Guilin Road, Shanghai 200234, China}

\author{Yifei Luo}
\affiliation{Department of Astronomy and Astrophysics, University of California, Santa Cruz, CA 95064, USA}

\author{Viraj Pandya}
\affiliation{Department of Astronomy and Astrophysics, University of California, Santa Cruz, CA 95064, USA}

\author{Rachel S.~Somerville}
\affiliation{Center for Computational Astrophysics, Flatiron Institute, 162 5th Avenue, New York, NY 10010, USA}
\affiliation{Department of Physics and Astronomy, Rutgers, The State University of New Jersey, 136 Frelinghuysen Rd, Piscataway, NJ 08854, USA}

\author{Henry C.~Ferguson}
\affiliation{Space Telescope Science Institute, 3700 San Martin Drive, Baltimore, MD 21218, USA}

\author{Susan Kassin}
\affiliation{Johns Hopkins University, Baltimore, MD 21218, USA}
\affiliation{Space Telescope Science Institute, 3700 San Martin Drive, Baltimore, MD 21218, USA}

\author{Anton M.~Koekemoer}
\affiliation{Space Telescope Science Institute, 3700 San Martin Drive, Baltimore, MD 21218, USA}

\author{Norman A.~Grogin}
\affiliation{Space Telescope Science Institute, 3700 San Martin Drive, Baltimore, MD 21218, USA}

\author{Audrey Galametz}
\affiliation{Max-Planck-Institut fur Extraterrestrische Physik, D-85748 Garching, Germany}

\author{P.~Santini}
\affiliation{INAF – Osservatorio Astronomico di Roma, via di Frascati 33, I-00040 Monte Porzio Catone, Roma, Italy}

\author{Hooshang Nayyeri}
\affiliation{Department of Physics and Astronomy, University of California, Irvine, CA 92697, USA}

\author{Mauro Stefanon}
\affiliation{Leiden Observatory, Leiden University, NL-2300 RA Leiden, Netherlands; Department of Physics and Astronomy, University of Missouri, Columbia, MO 65211, USA}

\author{Tomas Dahlen}
\affiliation{Space Telescope Science Institute, 3700 San Martin Drive, Baltimore, MD 21218, USA}

\author{Bahram Mobasher}
\affiliation{Department of Physics and Astronomy, University of California, Riverside, CA 92521, USA}

\author{Lei Hao}
\affiliation{Shanghai Astronomical Observatory, Chinese Academy of Sciences, Shanghai 200030, China}

\begin{abstract}
 This paper is one in a series that explores the importance of radius as a second parameter in galaxy evolution.  The topic investigated here is the relationship between star formation rate (SFR) and galaxy radius (\re) for main-sequence star-forming galaxies.  The key observational result is that, over a wide range of stellar mass and redshift in both CANDELS and SDSS, there is little trend between SFR and \re\ at fixed stellar mass. The Kennicutt-Schmidt law, or any similar density-related star formation law, then implies that smaller galaxies must have lower gas fractions than larger galaxies (at fixed \mstar), and this is supported by observations of gas in local star-forming galaxies. We investigate the implications by adopting the equilibrium ``bathtub'' model: the ISM gas mass is assumed to be constant over time and the net star formation rate is the difference between the accretion rate of gas onto the galaxy from the halo and the outflow rate due to winds. To match the observed null correlation between SFR and radius, the bathtub model requires that smaller galaxies at fixed mass have weaker galactic winds. Our hypothesis is that galaxies are a 2-parameter family whose properties are set mainly by halo mass and concentration. These determine the radius and gas accretion rate, which in turn predict how wind strength needs to vary with \re\ to keep SFR constant.
 
\end{abstract}


\section{Introduction}

Understanding how galaxies grow their stellar mass is one of the central questions in galaxy formation.  From observations, the global star formation rates of star-forming galaxies are observed to be well correlated with their stellar masses, a relation that has been termed the ``star-forming main-sequence'' (SFMS) \citep{Noeske-07,Elbaz-07,Elbaz-11,Daddi-07,Whitaker-12,Speagle-14}. Data show that this empirical relation has existed since $z$ $\geq$ 2 with a scatter of only 0.3 dex at fixed stellar mass \citep{Whitaker-12}.  Increasingly at late times, galaxies are found lying below the SFMS with star formation rates that are considerably lower than those on the star-forming ridgeline.  However, the present paper focuses on star-forming \emph{ridgeline galaxies}, which are clearly evident as a separate population at all redshifts \citep{Wuyts-11,Whitaker-17,Fang-18}.

\begin{figure*}
	\epsscale{1.0}
	\plotone{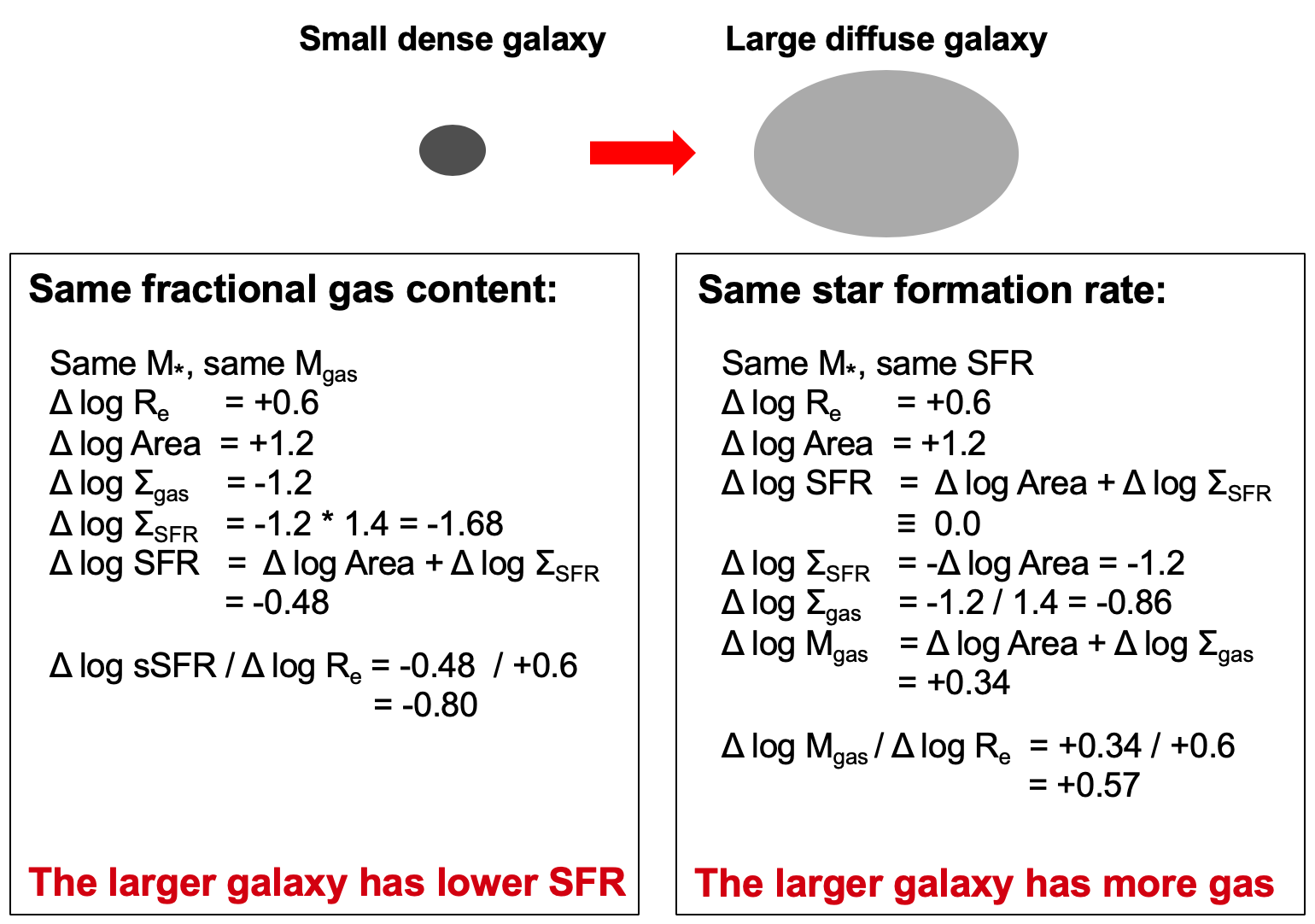}
	\caption{Prediction of SFR and gas differences by assuming constant gas fraction (left) or constant SFR (right). Following KS law, if large and small galaxies have the same gas fraction, it is predicted a slope of $-$0.80 in \delssfr\ vs. \dlRe\ plane. Otherwise, if they have the same SFR, it would have a slope of +0.57 in $\Delta$\,log\,$M_{\rm gas}$ vs. \dlRe\ plane.  
	\label{fig:cartoon}}
\end{figure*}

The SFMS is one of two major structural scaling relations for star-forming galaxies, the other being is the effective radius-stellar mass relation \citep[e.g.,][]{Shen-03,vanderWel-14}.  This relation also has scatter, and it is natural to consider whether residuals about the two relations are correlated.   There are at least two
reasons to think they might be.  The first stems from a simple model in which all galaxies obey the Kennicutt-Schmidt star formation law  \citep[KS law,][]{Kennicutt-98, Kennicutt-Evans-12}  and all galaxies at the same \mstar\ have the \emph{same gas fraction} regardless of \re.  One can then show (see prediction in Figure~\ref{fig:cartoon}) that large-\re\ galaxies would have lower total star formation rate than small ones, owing to the high exponent (1.4) in the KS law. Hence, a negative correlation is expected between residuals in sSFR and radius relative to the sSFR-mass and size-mass relations at fixed mass.

The second reason stems from a simple model for putting galaxies
into dark halos. This picture says that galaxy halos have two important structural parameters, \Mvir\ and concentration $C$ (or formation time), and that these parameters imprint themselves on galaxies to create the two-dimensional family of star-forming galaxies seen in \reff\ vs.~\mstar\ today. \Mvir\ maps onto \mstar, and concentration/formation-time plus \Rvir\ determines the baryonic radius of the galaxy forming within the halo \citep{Jiang-19}.  The latter effect arises because the centers of high-concentration halos collapse early when the universe is dense, thus forming a denser collapsed central object \citep{Wechsler-02}.  But high-concentration halos would accrete more slowly today (see Figure~\ref{fig:concentrationinfall}), and escape velocities would also be higher in denser galaxies, potentially producing weaker winds \citep{Dutton-vandenBosch-Dekel-10}.   Assuming the system follows the equilibrium bathtub model \citep[][see Section~\ref{subsec:equilibriumbathtub}]{Dekel-Mandelker-14}, halo mass accretion rate and wind mass-loading factor together determine the amount of gas available for star formation.  Hence a connection between galaxy radius residual and star formation residual is a possibility through their joint dependence on halo concentration/formation time.  

A number of theoretical papers are beginning to explore the effect of halo concentration on galaxy properties.  \citet{Dutton-vandenBosch-Dekel-10} calculated the effect of concentration\footnote{For brevity, we will henceforth use only the term ``concentration'' and omit ``formation time'', but implicitly we always mean the two together, as their effects are similar.} on wind strength using a semi-analytic model (SAM).  A companion paper to this one  \citep{Chen-20} posits a model in which black holes are more massive in higher concentration halos and considers how that would affect the structure of quenching galaxies. The impact of concentration is visible in the EAGLE simulations, where several papers have examined its effect on the stellar-mass-halo-mass relation \citep{Matthee-17,Kulier-19}, the star-forming main-sequence \citep{Matthee-Schaye-19}, and black hole mass and quenching \citep{Davies-19,Oppenheimer-19}.

This paper focuses on the impact of concentration on halo mass accretion rate, galaxy radii, winds, and star formation rates.   The launching point is the ``bathtub'' model \citep[e.g.,][]{Finlator-Dave-08,Bouche-10,Krumholz-Dekel-12, Dave-Finlator-Oppenheimer-12,Lilly-13,Forbes-14,Dekel-Mandelker-14,Schaye-15,Rodriguez-Puebla-16a}, sometimes also called the ``self-regulator'' model, in which each type of gas flow is represented by a \emph{single value} summed over the whole galaxy.  The foundation of the bathtub model is the Kennicutt-Schmidt law \citep{Kennicutt-98, Kennicutt-Evans-12}, in which SFR is a positive power of gas surface density.  Hence, if too much gas piles up, SFR increases, and the gas mass goes down.  Since star formation responds much more quickly than accretion onto halos changes \citep{Dutton-vandenBosch-Dekel-10,Dekel-13,Dekel-Mandelker-14}, the star formation rate continually adjusts itself to follow the accretion, and an equilibrium solution is reached in which the gas mass in the ISM is constant with time \citep{Dekel-Mandelker-14}.

Since new gas cannot add to the ISM, only two paths are available: make new stars or flow out as a wind.  The star formation rate is therefore really the difference between the accretion rate and the wind strength, and the ISM is simply a way-station through which gas passes on its way to making stars or wind.  From this point of view, the KS law should be read backwards: the galaxy needs some value of the SFR to be in equilibrium with the accretion rate and wind, and the ISM adjusts its density (according to the KS law) to make that happen. A more intuitive way to plot KS law would be to plot the required star formation rate on the $X$-axis as the independent variable and the resulting needed gas density on the $Y$-axis as the dependent variable. 

We can now ask what the bathtub model says about the effect of halo concentration on the star formation rate.  Higher concentration means higher infall at early times \citep{Wechsler-02}, and therefore, at fixed halo mass, less infall, and potentially lower SFR, at late times.  But higher early infall also makes a denser center, and the radius of the resulting galaxy will be smaller.  For the same stellar mass, such a galaxy will have a higher escape velocity, which might mean a weaker wind and more gas going into stars \citep{Dutton-vandenBosch-Dekel-10}.  Thus, whether the net star formation goes down due to decreased infall depends on whether that effect is balanced by increased star formation efficiency due to a weaker wind.

Which of these effects wins can be ascertained observationally by testing for a correlation between the radius residual \delsma\ and the star formation residual \delssfr.  We know for a fact that halo concentrations vary, and we show in Section~\ref{subsec:varyaccretion} that the expected effect on accretion from varying halo concentration is considerable.   If unopposed, the impact on the star formation rate should be obvious.  The observed null correlation thus sets a clear constraint on wind strengths that theoretical models must match. Indeed, in addition to the stellar-mass-halo-mass relation, this could be one of the tightest constraints on wind strength that we have.

The above logic suggests that studying \delssfr\ vs. \delsma\ might constrain winds.  Several works have previously examined such data in slices at fixed \mstar.  Across all galaxies, the broad trend is that galaxies \emph{well below} the SFMS are smaller than galaxies on the ridgeline, both locally \citep[e.g.][]{Shen-03,Omand-Balogh-Poggianti-14} and far away \citep{Wuyts-11,vanderWel-14}.  However, we are interested in the behavior of star-forming galaxies with \dlssfr\ $\ge$ $-$0.45 (our definition of the ridgeline in this paper). Within this range, \citet{Wuyts-11} found that SDSS galaxies above the SFMS are up to 0.3 dex smaller, but this trend shrank at intermediate redshifts and disappeared by $z$ = 2.0$-$2.5.   \citet{Fang-18} found no significant trend with size for CANDELS galaxies near and above the ridgeline, but galaxies 0.5 dex below the ridgeline were smaller.  \citet{Brennan-17} redid the \citet{Wuyts-11} study of SDSS galaxies taking care to eliminate galaxies with bad photometric fits, and Wuyts'  trend toward smaller galaxies above the SFMS nearly disappeared.  However, an L-shaped trend emerged at all redshifts whereby sizes at and above the ridgeline were flat with SFR while galaxies near the bottom of the ridgeline were 0.3$-$1.0 dex smaller, in agreement with \citet{Fang-18}.  \citet{Omand-Balogh-Poggianti-14} coded SDSS galaxies by SFR in the \re$-$\mstar\ diagram and saw no trend with radius for strongly star-forming galaxies but a decline in radius at very low star formation rates.  Finally, \citet{Whitaker-17} studied 3D-HST/CANDELS galaxies by stacking \emph{Spitzer} 24\um\ SFR values and found little trend with radius  except for a decline in SFR for very small galaxies at low redshift. 

  In summary, there appear to be two types of star-forming galaxies in these studies.  One type is strongly star forming near and above the peak of the ridgeline, and among them there seems to be little trend in SFR with radius or vice versa.  However, galaxies near the very bottom of the ridgeline appear to be smaller, the more so at lower redshifts.  Perhaps these objects are in transit to the green valley, where luminous radii shrink due to disk fading \citep{Fang-13}. These galaxies are seen in our sample as well, and they are mentioned in the Discussion.  On balance, though, the trend in SFR with radius for galaxies near and above the peak of the ridgeline is small.

 In this work, we start by investigating again the dependence between SFR and size for star-forming main-sequence galaxies. Compared to previous works, we implement some improvements.  First, we analyze CANDELS and SDSS in parallel, and all five CANDELS fields are used to maximize the sample. In the CANDELS sample, we use SFRs from dust-corrected NUV luminosities but compare those rates first to other 24\um\ values. The use of corrected NUV rates (in contrast to IR values) yields large samples down to 10$^9$\msun\ out to $z$ $\sim$ 2.5, and it also allows us to plot individual galaxies in the SFR-size plane without stacking. This latter point preserves the information in the 2-D distributions and lets us identify sub-populations, measure SFR as a function of \re\ and vice versa, and correlate dust absorption with location in SFR vs.~size.  In the SDSS sample, we use SFRs from \citet{Salim-18}, in which the SFRs are derived from SED fitting based on GALEX-SDSS-WISE photometry. In both samples, only face-on galaxies are used in order to minimize dust effects and biases in galaxy radii, and interacting and disturbed galaxies are also removed.  Centrals and satellites are studied separately in SDSS.  Previous studies tended to lump all masses together whereas our finer cuts reveal trends more clearly as a function of both time and mass.

 The upshot is to confirm more strongly the \emph{lack of any significant trend} between star formation rate and galaxy size among main-sequence ridgeline galaxies.  The KS law then predicts less total gas at fixed \mstar\ in smaller galaxies, which we confirm using measurements of \HI\ and H$_2$ in local galaxies.  This provides important independent validation that our star formation rate measurements are correct.  
  
We then interpret these results by adopting the bathtub model. Based on recent findings, we assume that small-radii galaxies sit in high-concentration halos \citep{Jiang-19}, and an N-body simulation is used to parametrize accretion rate vs.~halo concentration. The observed null trend in SFR vs. radius is then used to deduce the necessary change in wind strength vs.~galaxy radius needed to counteract the trend.  This is compared to the semi-analytic model of \citet{Dutton-vandenBosch-Dekel-10} and reasonable agreement is seen. In summary, if this chain of reasoning is correct, the primary conclusion is that the lack of an observed trend between galaxy radius and star formation rate implies that winds must be weaker in small galaxies at fixed stellar mass. A secondary conclusion is that radius is an important second parameter in galaxy evolution that may correlate with halo concentration, halo accretion rate, wind strength, and star formation history.

The paper is organized as follows. Section~\ref{sec:data} describes the data and sample selections. The main observational results on SFR vs.~radius are presented in Section~\ref{sec:result}. The resulting predictions for gas content vs.~galaxy radius are compared to local gas measurements in Section~\ref{sec:localgas}, where agreement is obtained.  Implications are discussed in Section~\ref{sec:discussion}.   Section~\ref{subsec:equilibriumbathtub} presents the basic bathtub model.  Section~\ref{subsec:varyaccretion} reviews evidence that galaxy radius depends on halo concentration and parametrizes how halo infall rate and therefore wind strength should vary vs.~galaxy radius.  This result is compared to data in the \citet{Dutton-vandenBosch-Dekel-10} SAM and in the EAGLE simulation.  Finally, Section~\ref{sec:conclusion} presents a summary and conclusions.  

Throughout this paper, we adopt a flat $\Lambda$CDM cosmology, with parameters $\Omega_{m}$ = 0.3, $\Omega_{\lambda}$ = 0.7, $H_{0}$ = 70 km s$^{-1}$ Mpc$^{-1}$. Values of \mstar\ and SFR are based on a \citet{Chabrier-03} initial mass function.  Occasionally we use the terms ``compact'' and ``diffuse'' to describe galaxy radii.  Compact simply means that the galaxy is smaller than average for its stellar mass, and diffuse means that it is larger than average.

\section{Data and Sample Selection}
\label{sec:data}

\subsection{Star-forming Galaxies in CANDELS}
\label{subsec:CANDELSsample}

\begin{figure*}
	\epsscale{1.2}
	\plotone{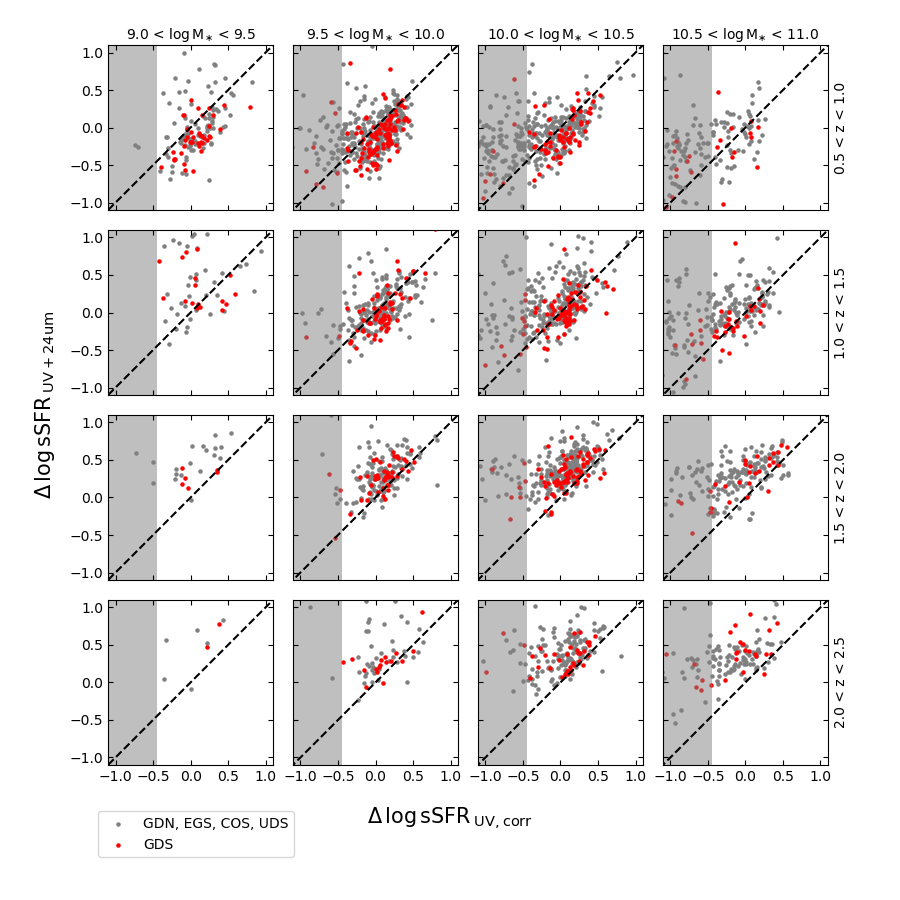}   
	\caption{Comparison of \dlssfruv\ used in this paper with \dlssfrIR\ based on the hybrid UV+IR SFR indicator described by \citet{Fang-18}. Face-on star-forming galaxies with good 24$\mu$m detections in all five CANDELS fields are shown.  Residuals in both sSFR$_{\rm UV,corr}$ and sSFR$_{\rm UV+24um}$ are calculated using identical main-sequence ridgelines, taken from \citet{Fang-18}. Green valley galaxies have \dlssfruv\ $<$ $-$0.45 dex and lie in the gray rectangles, while ridgeline galaxies populate the white areas. The dashed line represents the one-to-one relation. Clear correlations are visible in most of the ridgeline samples.  Zero-point offsets vary among the panels but do not disturb the \emph{relative} rankings within the SFMS ridgeline, which are used in this paper. \label{fig:comp_SFR}}
\end{figure*}

In this work we use all five fields from the CANDELS survey  \citep{Grogin-11, Koekemoer-11}, published in public catalogs by \citet{Galametz-13} for UDS, \citet{Guo-13} for GOODS-S, \citet{Barro-17} for GOODS-N, \citet{Nayyeri-17} for COSMOS, and \citet{Stefanon-17} for EGS\footnote{Public catalogs for all these fields are available in the $Rainbow$ database \citep{Barro-11}}. Rest-frame photometry and photometric redshifts are calculated using EAZY \citep{Brammer-vanDokkum-Coppi-08}. Spectroscopic redshifts are used if they are available (22\% of the sample).  The official CANDELS mass catalog includes results from 10 different SED fitting methods \citep[see table 1 in ][]{Santini-15}.  Most of them adopted BC03 \citep{Bruzual-Charlot-03} stellar templates and used minimizing $\chi^{2}$ to determine the best-fit. We use the median stellar mass because it averages the assumptions of different star formation histories. The median stellar mass is robust with a typical estimated error of $\sim$0.1 dex. 

Star formation rates, galaxy sizes, and their residuals are computed following the methods of \citet{Fang-18}. We briefly summarize them here and refer the reader to \citet{Fang-18} for more details.

The SFRs used in this work come from dust-corrected NUV luminosity (2800 \AA). In order to obtain a robust value of dust attenuation (in rest-frame V-band, \Av), the results from five different SED-fitting methods \citep[labeled as 2a, 2d, 12a, 13a, and 14a in][]{Santini-15} are combined and the median \Av\ is selected. These methods were chosen based on their common use of $\tau$-models and the Calzetti attenuation law. The typical formal error for the median \Av\ is $\sim$0.1 mag. The Calzetti dust attenuation at 2800 \AA, \Anuv, is 1.8\Av, which is used to correct the observed NUV flux. Corrected NUV luminosity is then converted to SFR using the calibration from \citet{Kennicutt-Evans-12}: \sfruvc[\msun\,yr$^{-1}$] = 2.59$\times$10$^{-10}$ \Luvc[\Lsun]. Although the methods are different, we have verified that our rates from NUV-corrected fluxes are in fact virtually identical to rates derived from full SED fitting. 

Since the goal of our study is to analyze the properties of galaxies above and below the SFMS, SFR measurements must be good enough to derive accurate SFR \emph{residuals}.  The formal errors of our SFRs are small, but there might be systematic errors, in part because of the use of $\tau$-models and/or the Calzetti law.  A separate study (Liu et al., in prep.) is testing SED-fitting methods on non-$\tau$ star formation histories, with encouraging results. In the meantime, it is desirable to have a separate set of SFRs to compare with. The so-called  ``hybrid'' method (\sfruvir), which adds together raw UV and IR rates, is thought to be the most reliable \citep{Kennicutt-09,Hao-11}. We calculate the quantity \sfruvir\ using the formula in \citet{Wuyts-11}, where $L_{\rm IR}$ is determined from 24\um\ data using the calibration of \citet{Rujopakarn-13}.  Data sources and details are given in \citet{Fang-18}.  
	
As explained in \citet{Fang-18}, the usual method of testing methods by simply plotting one SF indicator against another is not good enough to establish the accuracy of SFR \emph{residuals}.  We accordingly compute residuals using the SFMS ridgelines in different redshifts given in \citet{Fang-18} and compare them in Figure~\ref{fig:comp_SFR}.  Red points are from GOODS-S, which use deeper 24\um\ data, and the gray points are for all other fields.   This figure updates and extends a similar figure in \citet{Fang-18} to all five CANDELS fields.

Several conclusions emerge.  First, it is apparent that the IR data are highly incomplete at low mass and high redshift, and therefore it would be impossible to carry out the kind of study undertaken in this paper using IR data alone -- use of UV-optical SEDs is essential.  The second issue is evident in the shaded rectangles, which denote galaxies in the green valley below \dlssfr\ $< -0.45$ dex (our ridgeline boundary).  It is well known that SFRs for green valley galaxies are systematically overestimated by 24\um\ data according to several studies reviewed by \citet{Fang-18}, and the same trend is seen here. If these GV objects are set aside, clear if somewhat noisy correlations are visible in most of the panels.  The total scatter for the GOODS-S data (after removing outliers and zero point offsets) is 0.24 dex  \citep{Fang-18}, which, if assigned equally to both measures, implies an error of 0.17 dex in \dlssfruv.

The last point is the presence of systematic zero point offsets that tend to be negative at low redshift and positive at high redshift.  These are not a concern since our goal is the \emph{relative ranking} of objects within the SFMS, which is not disturbed by a zero-point shift.

The second important quantity used in this work is galaxy size defined by the half-light optical radii.  For this, we use the semi-major axes ($R_{\rm e, maj}$ = $a$) based on GALFIT fits to $H$-band images by \citet{vanderWel-14}.  $R_{\rm e, maj}$ is preferred to the circularized radius ($R_{\rm e, circ}$ = $\sqrt{ab}$) because it is a more stable indicator for inclined disks. Although $H$-band corresponds to different rest-frame wavelengths at different redshifts, \citet{Fang-18} estimated that, for SF galaxies between redshift $z$ = 1 and 2, the wavelength-dependent  $K$-correction to galaxy size is less than 10\%.  The residuals of $R_{\rm e, maj}$ are calculated using the mass-size relations from \citet{Fang-18} in different redshift bins.  Our use of residuals in bins of mass and redshift makes us insensitive to $K$-correction errors.

Finally we select only face-on galaxies with $b/a > 0.5$ in order to minimize the effects of dust on \dlssfruv\ and inclination on radii.  A summary of the CANDELS selection cuts is as follows:

\begin{enumerate}
    \item Apparent $H$-band magnitude $<$ 24.5. This is the limit suggested by \citet{vanderWel-14} for reliable GALFIT measurements.
    \item Redshift within 0.5 $< z <$ 2.5 and stellar mass within 9.0 $<$ log\,\mstar/\msun $<$ 11.0, to maximize the sample size. We omit log\,\mstar\ $>$ 11.0 and $z$ $<$ 0.5 because of few objects in those range.    
    \item PHOTFLAG = 0, CLASS\_STAR $<$ 0.9, and GALFIT flag = 0, to ensure reliable photometric measurements, no foreground stars, and good-quality GALFIT fits.  This eliminates merging galaxies and galaxies with peculiar morphologies.
    \item $b/a$ $>$ 0.5, to minimize dust extinction and radius uncertainties. $a$ and $b$ are the semi-major and semi-minor axes of the galaxy from GALFIT measurements.
    \item Location in the star-forming region of the UVJ diagram. Star-forming ridgeline galaxies must in addition have \dlssfruv\ $>$ $-$0.45 dex. Galaxies located in the SF region but with \dlssfruv\ $<$ $-$0.45 dex are retained but classified as green valley galaxies.
\end{enumerate}

According to the analysis in \citet{Fang-18}, these criteria include nearly all star-forming galaxies in most mass and redshift bins, but the completeness declines to less than 50\% for \mstar\ $<$ 10$^{9.5}$\msun\ and $z >$ 2.  See the discussion and figure 2 in \citet{Fang-18} on completeness.

\subsection{Star-forming Galaxies in SDSS}
\label{subsec:SDSSsample}

As a supplement to the high-redshift CANDELS data, we select normal star-forming galaxies from the SDSS DR7 catalog \citep{Abazajian-09}. Spectroscopic redshifts, stellar masses, and emission line measurements are obtained from the MPA/JHU value-added catalog \citep{Kauffmann-03a}.  

SFR and \Av\ are taken from \citet{Salim-18}, based on UV-optical SED fitting jointly with 22\um\ photometry. They compared their SFRs with other published catalogs and found good agreement for star-forming galaxies, which means that our results should not vary with different SFR indicators. The typical error in the star formation rate is about 0.1 dex \citep{Salim-16, Salim-18} . 

The SFMS ridgeline from \citet{Speagle-14} was adopted to calculate SFR residuals for the SDSS sample at fixed stellar mass (\dlssfr).  On average, we find that the mean sSFR in the Salim catalog is 0.14 dex higher than the ridgeline in Speagle. Since we are only concerned with relative sSFR, we adopt the slope of SFMS in \citet{Speagle-14} and shift the zero-point by 0.14 dex to match the SFR in \citet{Salim-16}.

Galaxy radii and \sersic\ indices are taken from the NYU Value-Added Galaxy Catalog \citep{Blanton-05}. Similar to $R_{\rm e, maj}$ in the CANDELS sample, we use the semi-major axis in $z$-band images to characterize galaxy size. We further calculate the size residuals according to the mass-size relation for star-forming galaxies in \citet{Shen-03}.

A summary of the criteria used to select the SDSS star-forming sample is as follows: 

\begin{enumerate}  
	\item Redshift in the range 0.02 $< z <$ 0.07, the apparent magnitude within 14 $ < r < $ 17.5,  and \mstar\ $>$ 10$^{9.0}$\msun.
    \item Single-\sersic\ index in the range 0.5 $<$ $n$ $<$ 6.  The upper limit excludes galaxies with bad fits.
    \item Merging galaxies are excluded using the classification from Galaxy Zoo ($P_{\rm MG}$ $<$ 0.1). (This is analogous to the good GALFIT flag that we required for CANDELS.) 
    \item $b/a$ $>$ 0.5. $a$ and $b$ are taken from the \citet{Blanton-05} catalog by single-\sersic\ fitting.
    \item Main-sequence membership using \dlssfr\ $>$ $-$0.45 dex after shifting the \citet{Speagle-14} zero-point by 0.14 dex (see above).  
\end{enumerate}

	Galaxies satisfying the above criteria comprise the full SDSS sample, which includes all galaxies on the main-sequence and is comparable to the CANDELS sample. In addition, to minimize possible environmental effects on the structure of the lower main-sequence and entrance to the green valley, we extracted a centrals-only subsample by matching to the group catalog of \citet{Yang-12} and requiring mass rank $M_{\rm rank}$ = 1. Finally, since another common way to select star-forming galaxies is by their emission lines, we constructed yet another subsample by requiring strong emission (S/N of [OIII]\,$\lambda$5007, H$\beta$, H$\alpha$, [NII]\,$\lambda$6584 $>$ 5) and location in the BPT diagram in the \HII\ region of \citet{Kauffmann-03b}.

\begin{figure*}
	\epsscale{1.1}
	\plotone{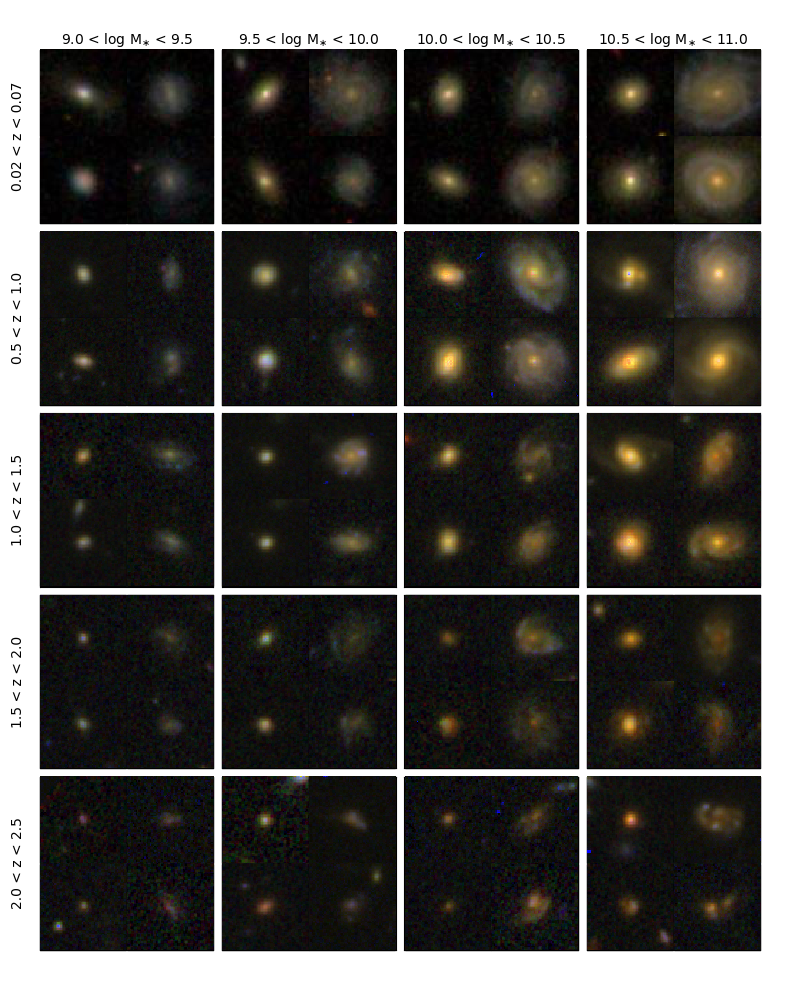}
	\caption{Color images of a random sample of CANDELS and SDSS star-forming galaxies demonstrating the difference between large and small galaxies. Local SDSS images come from $gri$ composites (top row); CANDELS images are generated from HST/ACS F814W, F125W and F160W. In each mass and redshift bin, the two small galaxies on the left are randomly selected from galaxies with \dlRe\ $<$ $-$0.2 dex, and the two large galaxies on the right are randomly selected from galaxies with \dlRe\ $>$ 0.2 dex. All images are scaled to span 30 kpc on a side.\label{fig:example}}
\end{figure*}

\section{Results}
\label{sec:result}

\subsection{No Trend Between SFR and Effective Radius}
\label{subsec:notrendbetweenSFRandR}

We turn now to correlations between SFR and radius.  Before continuing, we note that the RMS scatter in log\,\re\ at fixed \mstar\ is only about 0.25 dex \citep{vanderWel-14}, and it might be thought that measurement errors might mask real size differences.  To allay that concern,  Figure~\ref{fig:example} shows color images of large and small galaxies from both samples. In each mass and redshift bin, the two small galaxies on the left are randomly selected from galaxies with \dlRe\ $<$ $-$0.2 dex, and the two large galaxies on the right are randomly selected from galaxies with \dlRe\ $>$ 0.2 dex. The SDSS images at low redshift come from $gri$ composites while CANDELS images are generated from HST/ACS F814W, F125W, and F160W. Each image is presented at the same physical scale of 30 kpc on a side. It is seen that galaxies have distinctly different sizes and that \re\ does a good job of separating small galaxies from large ones. Since surface density is proportional to $r^{-2}$, the observed range of galaxy size results in a large change in surface density. A typical difference of $\pm$0.25 dex in radius would cause surface densities to differ by 1 whole dex.

\begin{figure*}
\epsscale{1.12}
\plotone{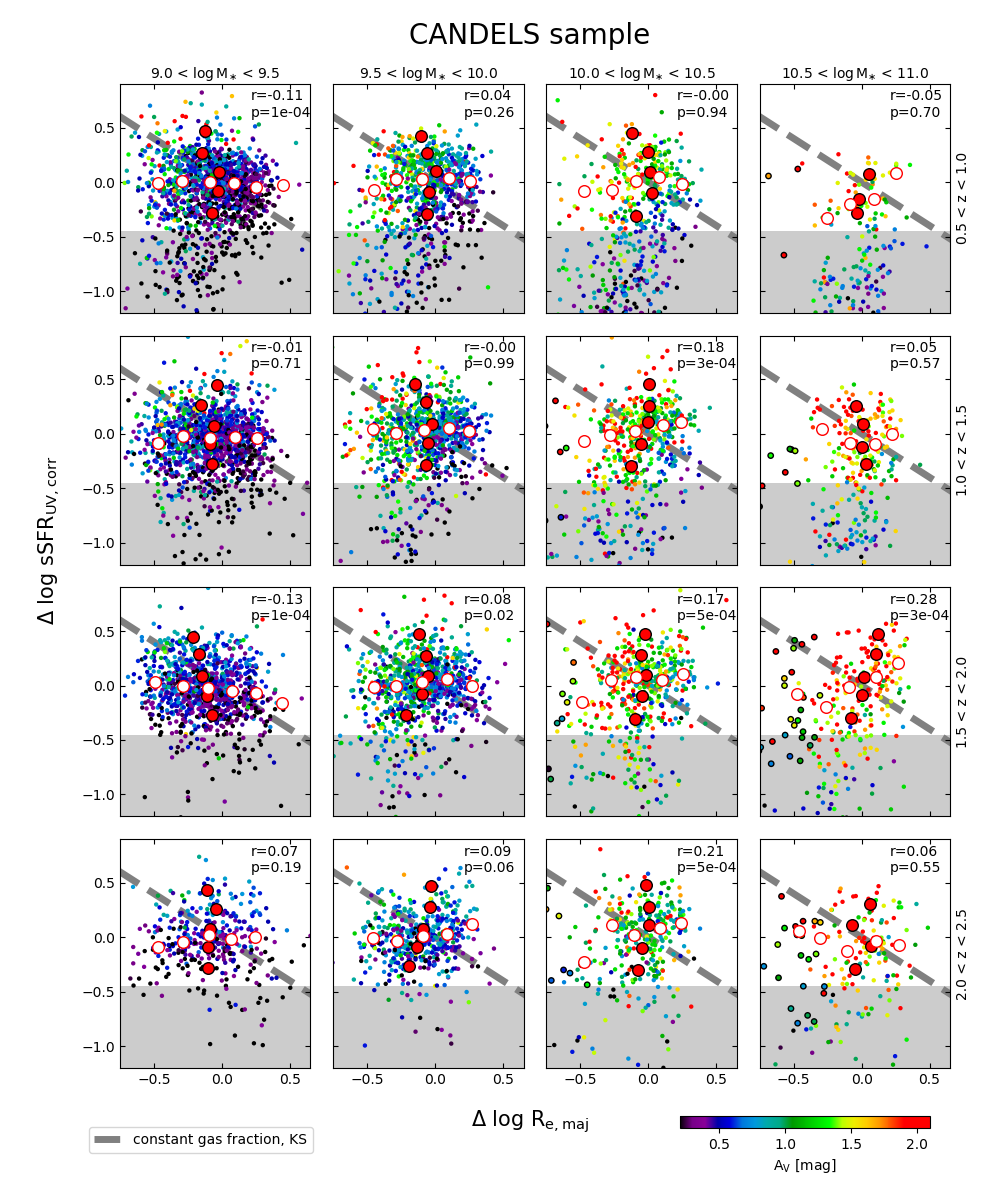}
\caption{Star formation rate residual \dlssfr\ vs.~radius residual \dlRe\ (semi-major axis) for face-on ($b/a >$ 0.5) CANDELS galaxies located in the star-forming region of the UVJ diagram. \dlssfr\ and \dlRe\ are the residuals from the star-forming main-sequence and the mass-size relations respectively \citep{Fang-18}. Points are color-coded by \Av\ from SED fitting.  Galaxies with \dlssfr\ more than 0.45 dex below the main-sequence ridgeline (in the shaded region) are classed as green valley galaxies and are excluded for calculating correlation coefficients. The solid red points are medians of \dlRe\ at fixed sSFR, the solid white points are medians of \delssfr\ at fixed \re. The Pearson correlation coefficient $r$ and the $p$-values are listed in the top-right corner of each panel.  The dashed lines are the predicted relations following the KS law under the assumption of constant gas fractions at fixed stellar mass. Their slopes are $-$0.80. Black circled points are galaxies that qualify as ``blue nuggets'' using the size-mass criterion of \citet{Barro-14}.  The data do not follow these predictions, indicating that smaller galaxies must have lower gas fractions.  Details are discussed in Section~\ref{subsec:notrendbetweenSFRandR}. \label{fig:dlssfr-dlsma} }
\end{figure*}

Figure~\ref{fig:dlssfr-dlsma} now plots the residuals \dlRe\ vs.~\dlssfr\ for CANDELS ridgeline galaxies divided into stellar mass and redshift bins. Points are color-coded by \Av\ obtained from SED fitting.  The trend predicted by the KS law under the assumption of constant gas fraction (see below) is the gray line in each panel.  Ridgeline galaxies with \dlssfr\ $\ge$ $-$0.45 dex are in the white areas; green valley (GV) galaxies are in the shaded regions. Compared to ridgeline galaxies, the latter tend to have smaller radii and lower \Av.   However, for ridgeline galaxies only, there is no strong trend for \dlssfr\ to follow the gray lines predicted by the KS relation for constant gas fraction.

The lack of any significant correlation is confirmed by the Pearson correlation coefficient $r$ and the $p$-values, which are listed in the top-right corner of each panel. The coefficients $r$ are close to zero in most panels, indicating no correlation between the two variables. The $p$-value indicates the probability of an uncorrelated system producing relations that have a Pearson correlation as large as the one computed from these datasets. In all panels, the correlation coefficients are low ($r<$ 0.28). For low-mass galaxies with \mstar\ $<$ 10$^{10}$ \msun, the signs of $r$ vary randomly and $p$-values are mostly not significant ($>$ 0.05). At higher masses \mstar\ $>$ 10$^{10}$ \msun\ there is a slight trend to see positive slopes from the white points, and half of the $p$-values are significant. In any case, there is no sign of the systematically negative slopes that are predicted by the KS relation.
	
To gain further insight, we plot medians of \dlssfr\ at fixed \dlRe\ (white points) and medians of \dlRe\ at fixed \dlssfr\ (red points). In all panels, the two sets of lines outlined by these points are orthogonal or nearly so, which is another classic signature of little or no correlation. It is interesting to note that the mean slopes indicated by the white points above  \mstar\ $>$ 10$^{10}$ \msun\ are weakly positive. This trend follows the numbers of galaxies near the bottom of the ridgeline and in the green valley, which are also increasing with mass.  Since these galaxies have both small radii and low SFR, their presence tends to create a positive slope.  We note that the trend for galaxies well below the ridgeline to have smaller radii was also seen by \citet{Brennan-17} for CANDELS, but we have now shown that this trend is correlated with the number of green valley galaxies. An inference might be that a significant fraction of star-forming galaxies below the ridgeline at high mass is actually en route to the green valley, i.e., that they are not bobbing temporarily below the ridgeline and are about to return. However, the bigger picture is that even the largest slope of $\sim$0.2 is small, amounting to a total variation of only 0.1 dex (=25\%) from small ($-$0.25 dex) to large (+0.25 dex) galaxies. In addition, by comparing the slopes in different redshift bins, we do not see significant evolutionary trends with time.

We referred above to the prediction of the KS law (gray line), which says that, if gas fractions are all the same for galaxies in a given stellar mass bin, large, diffuse galaxies should have lower SFRs than compact, dense ones.  The prediction is computed in the cartoon in Figure~\ref{fig:cartoon}, which compares two galaxies of the same stellar mass, one large and one small.  The left panel describes the situation with identical gas fractions.  Spreading the same amount of gas over a wider area reduces its ability to form stars owing to the exponent 1.4 in the KS law, which exceeds unity.  The predicted slope is $-$0.80 for \dlssfr\ vs. \dlRe, which is shown as the gray lines in Figure~\ref{fig:dlssfr-dlsma}.  

For comparison, the case of constant star formation rate is shown in the right panel of Figure~\ref{fig:cartoon}, where it is shown that the larger galaxy must have more gas in order to make the same amount of stars.  The predicted slope in that case is +0.57 for $\Delta$\,log\,$M_{\rm gas}$ vs. \dlRe\ (we show this trend in Figure~\ref{fig:dlfgas}).

A new version of the KS law has recently been proposed called the extended Kennicutt-Schmidt (eKS) law, \sigsfr\ $\propto$ \siggas \sigstar$^{0.5}$ \citep{Shi-11}, which takes stellar surface density into account and improves the relation for low-surface-brightness galaxies \citep{Shi-18}.  Use the method of Figure~\ref{fig:cartoon}, the predicted slope for eKS is $-$1.00 for \dlssfr\ vs. \dlRe. Both predicted relations are far steeper than the data.  Hence, we conclude that, if these galaxies obey the  KS law or the eKS law, larger galaxies must have larger gas fractions.  This is confirmed for local galaxies in Section~\ref{sec:localgas} below.

Figure~\ref{fig:dlssfr-dlsma} is an opportunity to show the locations of the compact star-forming galaxies identified by \citet{Barro-14} in relation to other objects at the same redshift.   These so-called ``blue nugget'' galaxies were identified as having particularly small sizes compared to typical main-sequence ridgeline objects. We select them using the same definition log(\mstar/$R_{e}^{1.5}$) $>$ 10.45 \msun\ kpc$^{-1.5}$ used by \citet{Barro-14} and plot them with black circles in Figure~\ref{fig:dlssfr-dlsma}. Most compact star-forming galaxies in our sample appear at redshifts \redshift\ $>$ 1.5 and either lie on the main-sequence or below it, consistent with the scenario that blue nuggets are in the process of quenching and will soon evolve into red nuggets by \redshift\ = 1.5 \citep{Barro-14}.

\begin{figure*}
\epsscale{1.0}
\includegraphics[width=\textwidth]{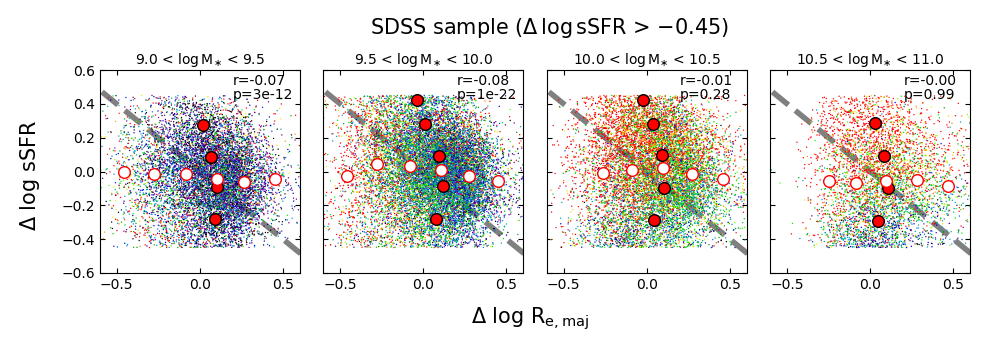}
\includegraphics[width=\textwidth]{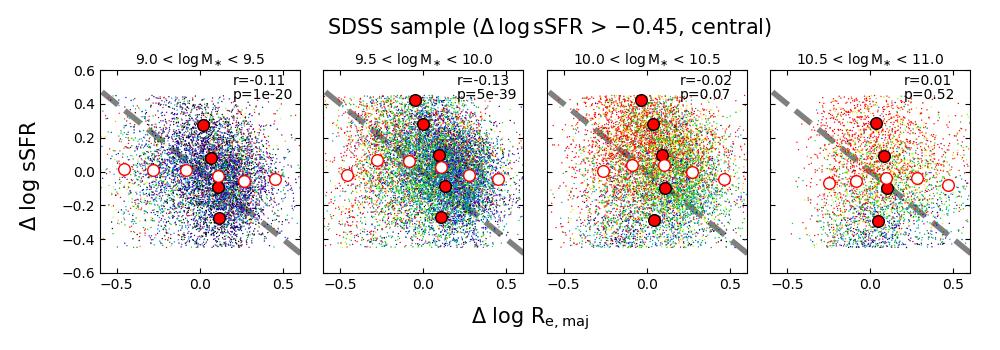}
\includegraphics[width=\textwidth]{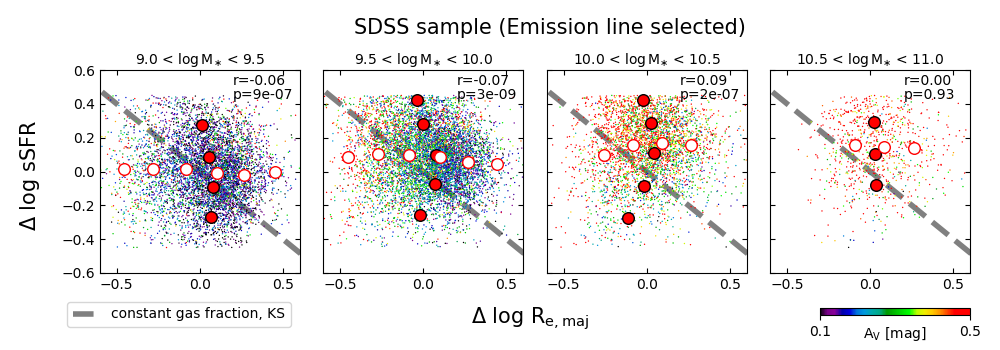}
\caption{\dlssfr\ versus \dlRe\ for the SDSS sample. Galaxies have $z$ = 0.02$-$0.07 and are face-on.  Top row: All galaxies.  Middle row: Central galaxies only.  Bottom row: Emission line sample (see Section \ref{subsec:SDSSsample} for sample descriptions). Points are colored by \Av\ from UV-optical SED fitting. The solid red points are medians of \dlRe\ at fixed sSFR, the solid white points are medians of \dlssfr\ at fixed \re. The Pearson correlation coefficients $r$ and $p$-values are listed in the top-right corner of each panel.  The dashed lines are the predictions from the KS law by assuming constant gas fraction. Overall, these samples also show little trend in SFR vs.~size, in agreement with the CANDELS results in Figure~\ref{fig:dlssfr-dlsma}. Details are discussed in Section \ref{subsec:notrendbetweenSFRandR}.  \label{fig:dlssfr-dlsma-sdss}}
\end{figure*}

\begin{figure*}
	\epsscale{1.0}
	\plotone{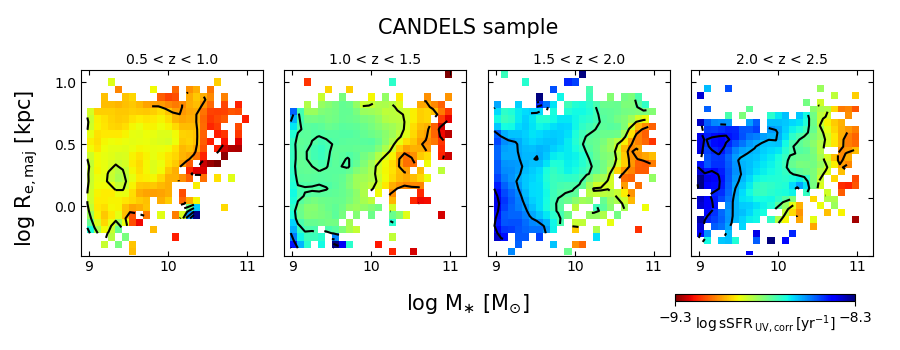}
	\caption{The mass-size relation for star-forming galaxies in different redshift bins in CANDELS. Stellar mass and galaxy radius are binned and colored by the median log\,sSFR in each pixel.  The black contours indicate the log\,sSFR level in steps of 0.25 dex. The contours are roughly vertical but sightly tilted, which is consistent with the small positive or negative slopes seen in Figure~\ref{fig:dlssfr-dlsma}. \label{fig:lmass_lSMA_mosaic}}
\end{figure*}

\begin{figure*}
	\epsscale{0.5}
	\centering
	\includegraphics[width=0.3\textwidth]{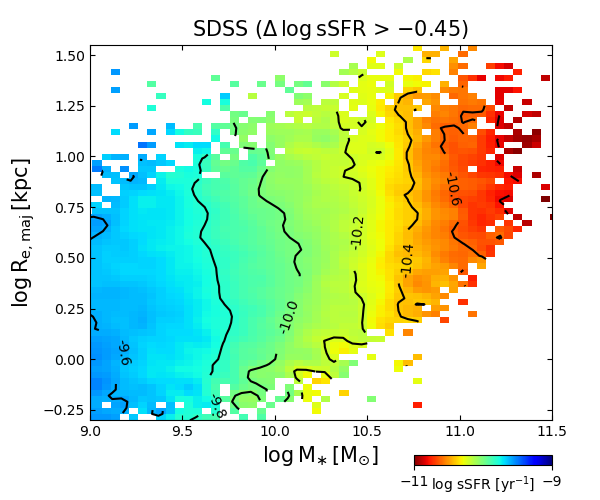}
	\includegraphics[width=0.3\textwidth]{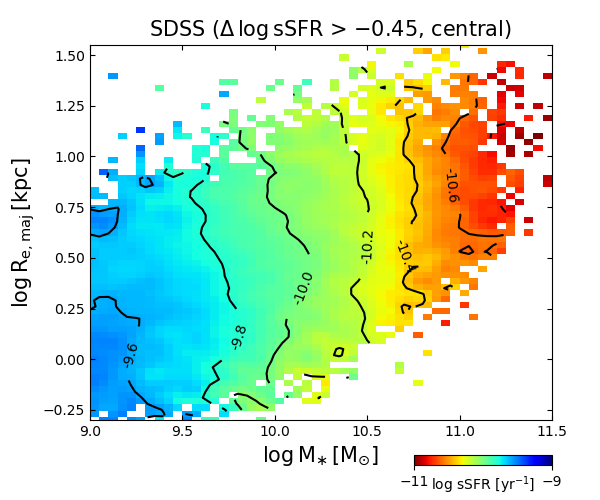}
	\includegraphics[width=0.3\textwidth]{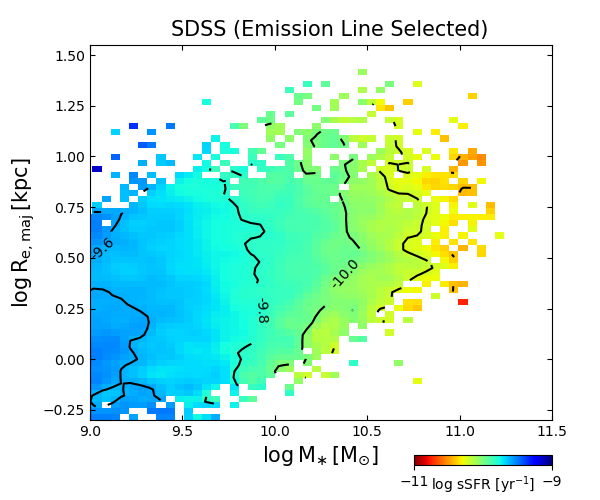}
	\caption{The mass-size relation for star-forming galaxies in SDSS samples.  Left: all galaxies; middle: central galaxies only; right: emission-selected sample. Stellar mass and galaxy size are binned and colored by the median log\,sSFR in each mosaic. The black contours indicate the log\,sSFR level in steps of 0.2 dex.  As for CANDELS in Figure~\ref{fig:lmass_lSMA_mosaic}, the contours are basically vertical but slightly tilted, consistent with the small positive slopes seen in Figure~\ref{fig:dlssfr-dlsma-sdss}.   \label{fig:lmass_lSMA_mosaic_sdss}}
\end{figure*}

Figure~\ref{fig:dlssfr-dlsma-sdss} shows \dlRe\ vs.~\dlssfr\ for the SDSS sample.   The first row shows the full sample, which follows the same selection criteria as CANDELS. The second row shows central galaxies only, while the third row selects strongly star-forming galaxies using the emission-line criterion (see Section~\ref{subsec:SDSSsample}).  The correlation coefficient $r$ and $p$-values are calculated for each panel as in Figure~\ref{fig:dlssfr-dlsma}. The medians of both $X$ and $Y$ directions are shown in each panel.  The predicted slopes according to the KS law are the dashed lines. 

As in Figure~\ref{fig:dlssfr-dlsma} for CANDELS, the correlation coefficients are low, the medians in $X$ and $Y$ directions are quite orthogonal, and the choice of sample also has little effect. The second row shows central galaxies only. Removing the satellites appears to have removed some of the small galaxies below the ridgeline at low masses, and the star-forming sample looks a bit cleaner.  However, the correlation coefficients and the slopes in $X$ and $Y$ directions are basically unaffected. The emission-selected sample in the third row has a tail of low-SFR galaxies with small sizes. This population resembles the similar population in CANDELS and seems stronger in the emission sample than the other samples. As before, however, the correlation coefficients are all small.

In conclusion, the result from SDSS agrees with CANDELS in showing that smaller star-forming ridegline galaxies must contain less gas at fixed \mstar\ if galaxies obey the KS or eKS star-forming laws.

We note in passing that the data points are colored by \Av\ in Figures~\ref{fig:dlssfr-dlsma} and \ref{fig:dlssfr-dlsma-sdss}. Even though both samples are deliberately restricted to face-on galaxies with $b/a>0.5$ in order to minimize dust effects, nevertheless two trends are evident. The stronger is that more compact galaxies have higher \Av\ than larger galaxies (this effect looks more prominent in SDSS, but note the compressed color range compared to CANDELS -- CANDELS is just noisier).  On the face of it, this is puzzling because we have just shown that more compact galaxies have \emph{less} gas, so why do they have higher \Av?  The answer is that \Av\ varies as the \emph{surface density} of gas, not the gas fraction.  Compact galaxies evidently produce larger \Av\ on account of their smaller area even with less total gas.   The second trend is that galaxies with high \dlssfr\ have higher \Av\ at fixed \dlRe.  This trend is plausible, since higher SFR at fixed size implies more gas, and thus more dust. Both trends will be explored in future papers.

Figures~\ref{fig:lmass_lSMA_mosaic} and \ref{fig:lmass_lSMA_mosaic_sdss} summarize the preceding data by plotting log\,sSFR as a function of position in the mass-size diagrams. Stellar mass and galaxy size are binned and colored by the median log\,sSFR in each pixel. The CANDELS sample in Figure~\ref{fig:dlssfr-dlsma} is a bit noisy and the contours do not vary smoothly, probably due to the smaller sample size or limitation of signal-to-noise ratio. Overall, however, they are roughly vertical but are slightly tilted at intermediate mass, consistent with the small positive or negative slopes in Figure~\ref{fig:dlssfr-dlsma}. In the SDSS sample (Figure~\ref{fig:lmass_lSMA_mosaic_sdss}), both the full and central samples show nearly vertical contours, but the emission-selected contours are more tilted, consistent with the trend for this sample in Figure~\ref{fig:dlssfr-dlsma-sdss}. 

To summarize, we have compared \dlssfr\ vs.~\dlRe\ for star-forming ridgeline galaxies over a wide range of stellar mass and redshift. Neither CANDELS nor SDSS shows a large trend in star formation rate vs.~galaxy radius at fixed stellar mass. This result does not depend on the SFR indicator used, nor does it appear to vary much with sample selection (in SDSS). The assumption of constant gas fraction at fixed stellar mass would predict a large negative trend between \dlssfr\ and \dlRe, which does not appear in the real data. If galaxies obey a density-dependent star-forming law like the KS law or its relatives, these results indicate that more compact galaxies have lower gas fractions.

\subsection{Comparison with Previous Work}
\label{subsec:previouswork}

We return now to the discussion of previous work that was initiated in the Introduction. The basic question is whether the properties of galaxies are correlated with their position above or below the star-forming main-sequence.  Our approach here has been to measure the correlation between main-sequence star-forming residual \dlssfr\ and radius residuals \dlRe. This is the same approach used by \citet{Whitaker-17} for CANDELS galaxies and by \citet{Omand-Balogh-Poggianti-14} for SDSS galaxies.  These papers also found no significant correlation if the sample is restricted to ridgeline galaxies, and we agree.

An alternative approach is to use the main-sequence residual \dlssfr\ as the basic variable and look for trends vs.~that.  This is the approach used by \citet{Wuyts-11} for SDSS and CANDELS galaxies and by \citet{Brennan-17} for GAMA and CANDELS galaxies.  As is well known, if there is significant scatter between two quantities $X$ and $Y$, the median of $X$ on $Y$ can behave differently from the median of $Y$ on $X$.  To facilitate comparison with \citet{Wuyts-11} and \citet{Brennan-17}, Figures~\ref{fig:dlssfr-dlsma} and \ref{fig:dlssfr-dlsma-sdss} cut the samples horizontally in slices of \dlssfr\ and show the median value in each slice (red circles).  The SDSS points reproduce closely the trend found by \citet{Brennan-17} using GAMA galaxies, showing the largest value \dlssfr\ on the ridgeline and declines amounting to $\sim$0.2 dex above and below it.  This also agrees with Wuyts' analysis of SDSS although the trends there were slightly larger. In CANDELS, none of the three works reports any significant trend in \dlRe\ across the main-sequence, all trends both above and below the SFMS being $\leq0.1$ dex. The data in Figure~\ref{fig:dlssfr-dlsma}, though noisy, agree with this.  However, galaxies well below the ridgeline but still with \dlssfr\ $> -0.45$ dex appear to be a little smaller in all works, and we have wondered whether this population is slowly quenching and moving towards the green valley.

In summary, the lack of any significant trend between \dlRe\ and \dlssfr\ for star-forming ridgeline galaxies is now well established from several different studies using different samples of galaxies at different redshifts.  Our study has divided galaxies by mass and redshift and plotted each mass-redshift bin as individual points.  This has revealed a probable population of compact galaxies below the SFMS that may be en route to the green valley and shown that this population is stronger at high masses.  We return to this population briefly in Section~\ref{sec:discussion}.  Our use of central galaxies and exclusion of mergers and galaxies with bad GALFIT fits has also removed any concern that improper sample selection might have colored previous results.  Finally, color-coding by \Av\ has revealed systematic trends for stronger reddening in galaxies above the SFMS and in galaxies with smaller radii.  These trends will be followed up in future papers.

\section{Confirmation Using Local Gas Measurements}\label{sec:localgas}

As noted, these results imply that the total gas fraction (H$_2$ + \HI) must be lower in small galaxies at fixed stellar mass.  It would be good to have direct confirmation of this, but observations of gas fractions are difficult at high redshift.   Indirect support comes from observations of H$_2$, as summarized by \citet{Tacconi-18}.  They find that the depletion time of molecular gas (\MHmol/\mdots) does not vary with \re\ on the main-sequence. Since SFR also does not vary, this means that on average all galaxies of a fixed mass but different radii must have the \emph{same} mass of H$_2$.  The KS and eKS laws then predict that the ratio $M_{\rm gas}$/\MHmol\ should be lower in smaller galaxies since their surface densities are higher, which favors conversion of \HI\ to  H$_2$.  Hence, total gas should be smaller.  We therefore obtain consistency with our results, but only by invoking the star formation laws.  It would be good to obtain confirmation for at least some populations of galaxies without appealing to those laws.

\begin{figure}
\epsscale{1.2}
\plotone{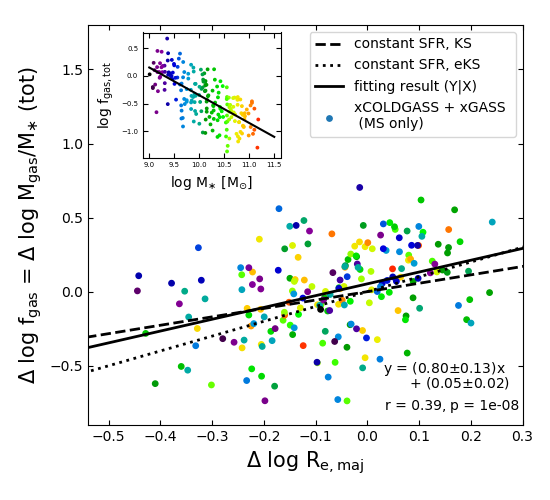}
\caption{Gas fraction vs.~galaxy size for star-forming galaxies based on data from the xCOLD$\,$GASS and xGASS surveys. Only galaxies on the star formation main-sequence are shown. Points are colored by stellar mass. A linear least-squares relation is fitted to gas fraction vs.~stellar mass in the upper left corner, and residuals relative to this relation are plotted vs.~galaxy-size residual in the main panel. The fitting result between two residuals is shown in the lower left corner, as well as the correlation coefficient. The dashed line is the prediction from the KS law, which has slope 0.57 (panel $b$ of Figure~\ref{fig:cartoon}). The dotted line is the prediction from the eKS law, with a slope of 1.00.  Overall, the data agree well with the predicted trend that smaller galaxies should have less gas at fixed stellar mass. The variation from small to large galaxies is a factor of 2-3.\label{fig:dlfgas}}
\end{figure}

This is possible at low redshift using measurements of local galaxies from the xCOLD$\,$GASS and xGASS surveys \citep{Saintonge-17, Catinella-18}. xCOLD$\,$GASS measured molecular hydrogen in a representative sample of 532 SDSS-selected galaxies with \mstar\ $> 10^{9}$\msun\ using CO(1-0) on the IRAM-30m telescope.  xGASS measured neutral hydrogen in 1200 galaxies in the same mass range using the Arecibo telescope. We match the xCOLD$\,$GASS and xGASS samples and only select galaxies on the main-sequence using \dlssfr\ $>$ $-$0.45 dex. Figure~\ref{fig:dlfgas} plots the results. The upper-left inset shows the basic data of \fgas\ vs. \mstar. Gas fraction is seen to decline smoothly as a function of stellar mass and is well fit by the linear least-squares fit of \fgas\ on \mstar\  (black line).  Residuals relative to this line are plotted vs.~\dlRe\ in the main panel.  Points are color-coded by stellar mass, and no systematic departures with mass are seen.   A least-squares fit of $Y$ on $X$ is shown by the solid line. The dashed line (a slope of 0.57) is the prediction from the SFR surface density (0.5$\times$SFR/$\pi$\re$^2$) assuming the KS law. Assuming the eKS law have a slightly steeper slope of 1.00.  The data are consistent overall with the prediction that smaller galaxies have less gas. From small to large galaxies, gas fractions vary on average by a factor of 2-3. 

We conclude by noting that the correlation in Figure~\ref{fig:dlfgas} is just the first step. To really test the KS law requires knowing the {\it gas surface density}, which we have implicitly estimated by assuming that gas radii are proportional to optical radii \re. In fact, this ratio varies from galaxy to galaxy \citep[see the range of $R_{\rm HI}$/$R_{90}$ in][]{Wang-20}. A second correction is needed for the fact that sSFR varies randomly from galaxy to galaxy within a given \mstar\ slice (i.e., scatter about the SFMS). Neither of these corrections has been made in Figure~\ref{fig:dlfgas}, which is why the scatter is so large. We have verified in work in progress that making both corrections significantly reduces the scatter in Figure~\ref{fig:dlfgas} but the slope remain unchanged, as expected from the work of \citet{Bruzzone-98}, which says that a least-squares fit of $Y$ on $X$ is a good estimate for the true slope when the error on $Y$ is much larger than the error in $X$. The large errors $Y$ are what justifies the slope estimate in Figure~\ref{fig:dlfgas}. The fully corrected relation will be presented in a future paper.

\section{Discussion}
\label{sec:discussion}

Our major observational result is that the star formation rate does not depend significantly on galaxy radius at fixed stellar mass for galaxies on the star-forming ridgeline.  This is true at all masses \mstar\ $>$ 10$^{9.0}$\msun\ and redshifts $z <$ 2.5.  Density-based star formation laws like the KS law or the extended KS law then predict that compact galaxies should have smaller total gas fractions than diffuse galaxies because of their higher gas densities. High densities in turn mean higher star formation rates because the exponent in the KS and related laws is greater than 1.0 \citep[1.4$-$1.5 in various versions of the KS law, see][]{Kennicutt-98}.   If the power is unity, it does not matter how the gas is distributed, and the star formation efficiency is the same for all gas distributions.  This is the case for the molecular law \citep[as reviewed in ][]{Kennicutt-Evans-12}, which is consistent with a model in which H$_2$ is located in individual clouds and the global distribution of those clouds does not matter.  H$_2$ mass at fixed \mstar\ should therefore be constant with size, and the smaller total gas masses in smaller galaxies are due to more efficient conversion of \HI\ to H$_2$.

In summary, total gas fraction should be lower in compact star-forming galaxies, but molecular gas content should be the same.  Similar conclusions were reached by \citet{Popping-15}, who studied CANDELS galaxies.  They used similar star formation rates to our values based on UV-optical SED fitting, but their conversion from star formation rate to gas density was more elaborate, taking mid-plane gas pressure into account.  Nevertheless, the two approaches are fundamentally comparable, and similar predictions emerge.

\subsection{The Equilibrium Bathtub Model}\label{subsec:equilibriumbathtub}

We turn to interpret these results by assuming the equilibrium bathtub model.  The basic equation for the model can be written \citep{Dekel-Mandelker-14, Rodriguez-Puebla-16a}:
\begin{equation}
    \dot{M}_{\rm in} = \dot{M}_{\ast} + \dot{M}_{\rm out} + \dot{M}_{\rm ISM} \\
    = (1+\eta)\dot{M}_{\ast} + \dot{M}_{\rm ISM},
\end{equation}
or,
\begin{equation}\label{eq:equilibriumbathtub}
    \dot{M}_{\rm ISM} = \dot{M}_{\rm in} - (1+\eta)\dot{M}_{\ast} = 0.
\end{equation}
where $\dot{M}_{\rm in}$ is the accretion rate of pristine gas into the halo, \mdots\ is the star formation rate, $\dot{M}_{\rm out}$ is the outflow rate, and $\dot{M}_{\rm ISM}$ is the rate of mass accumulation in the ISM. Using the mass-loading factor $\eta$ for the wind results in the second line.   Here we have assumed that all gas that falls into the halo finds its way soon into the galaxy, i.e., that gas is not accumulating in the halo.  We have also assumed that no wind gas falls back in, i.e., that it either escapes the halo or is inert.    \citet{Dekel-Mandelker-14} showed that the equilibrium solution under these circumstances is $\dot{M}_{\rm ISM} = 0$, which is reached asymptotically over time.  This is explained by the feedback in the sign of \mdots, which varies negatively with ${M}_{\rm ISM}$.  Errors in ${M}_{\rm ISM} $ are therefore self-correcting, and the star formation rate is self regulating.  The equilibrium  solution is obtained provided the response time for changes in \mdots\ is short compared to variations $\dot{M}_{\rm in}$.  \citet{Dekel-Mandelker-14} adopt for $t_{\ast}$ the local crossing time in the galaxy, or $R_d/V_d$, while $t_{\rm infall}$ is the crossing time of the halo, or $R_{\rm vir}/V_{\rm vir}$, which is $\sim$10 times longer.  The needed inequality is therefore satisfied.

The next step notes that, if $\dot{M}_{\rm ISM} = 0$, then $(1+\eta)\dot{M}_{\ast} = \dot{M}_{\rm in}$, i.e., that the star formation rate is proportional to the halo gas accretion rate. Eq.~\ref{eq:equilibriumbathtub} says that there are only two ways of increasing the star formation rate: larger halo accretion rate (\mdotin) or a weaker wind (smaller $\eta$).  In particular, increasing the gas surface density (by, say, reducing galaxy radius) or increasing the local star formation efficiency (by, say, raising the coefficient in the KS law) does not make more stars $--$ it cannot because the total gas supply is limited.  The only consequence of raising the local star formation efficiency is to reduce the mass of the ISM that it takes to support the same star formation rate.  In other words, the gas density is adjusting itself to accommodate the halo accretion rate, and the proper way to read the KS law is backwards, from $Y$-axis to $X$-axis, as suggested in the Introduction.

\subsection{The Bathtub Model with Variable Concentration}\label{subsec:varyaccretion}

\begin{figure*}
\epsscale{1.2}
\plotone{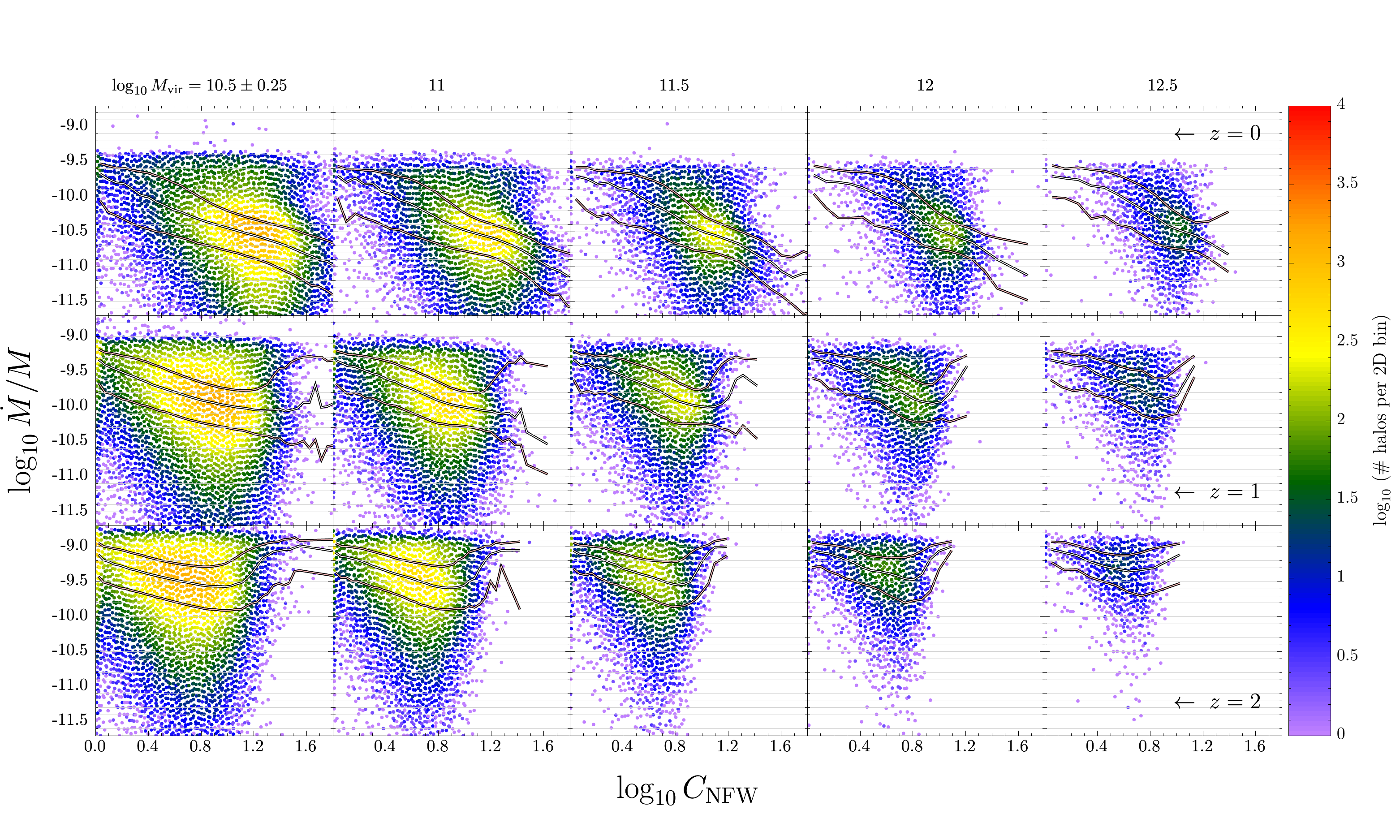}
\caption{Bivariate distributions of specific halo mass accretion rate vs.~halo concentration parameter $C_{\rm NFW}$ \citep{Navarro-96} binned by mass and redshift.  The lines indicate mean $\langle \dot{M}_{\rm in}/M_{\rm vir} \rangle$ at each $C_{\rm NFW}$ and the $\pm$1-$\sigma$ contours.  An average fit to all bins is $\dot{M}_{\rm in}/M_{vir}  \sim (C/\langle C \rangle)^{-0.5}$. High-concentration halos accrete less at any epoch because they accrete a greater fraction of their mass earlier and less later.  \label{fig:concentrationinfall}}
\end{figure*}

We now use the equilibrium bathtub model to see how galaxies respond to varying halo concentration.  The focus on concentration is motivated by an analysis of galaxy radii from the VELA and NIHAO simulations by \citet{Jiang-19}, who find that 
\begin{equation}\label{eq:radiiconcentration}
    R_{\rm e} = 0.02 R_{\rm vir} (C/10)^{-0.7}
\end{equation}
where \re\ is the 3-D mass-weighted half-mass radius and $C$ is halo concentration.  The trend with $C$ reflects the fact that galaxies in higher-concentration halos are smaller because a larger fraction of their mass is accreted early, making a smaller and denser galaxy.  As a result halos with higher concentration at fixed $M_{\rm vir}$ and fixed epoch have lower \mdotin.  This is illustrated in Figure~\ref{fig:concentrationinfall}, which plots the bivariate distributions of specific halo accretion rates, $\dot{M}_{\rm in}/M_{\rm vir}$, vs.~halo mass from the Rockstar halo catalog \citep{behroozi-19} based on the Bolshoi-Planck dark-matter simulation \citep{Klypin-16,Rodriguez-Puebla-16b,Lee-17}.  The data are in panels binned by halo mass and redshift, and the lines show the mean and $\pm$1-$\sigma$ contours.  A good fit to the average instantaneous specific accretion rate is:
\begin{equation}\label{eq:accretionconcentration}
      \dot{M}_{\rm in}/M_{\rm vir}  \sim (C/\langle C \rangle)^{-0.5},
\end{equation}
where $C$ is the instantaneous concentration.  The negative trend reflects the fact that high-concentration halos accreted more of their mass early and thus accrete less mass later. Putting the two equations together yields the prediction:
\begin{equation}\label{eq:accretionradius}
      \dot{M}_{\rm in}/M_{\rm vir}  \sim R_{\rm e}^{0.7}.
\end{equation}

Hence, the halo mass accretion rate should be lower in compact galaxies.  At fixed $M_{\rm vir}$, \mstar\ and $\eta$, Eqs.~\ref{eq:equilibriumbathtub} and \ref{eq:accretionradius} imply \dlssfr/\dlRe = +0.7.  In the context of Figures~\ref{fig:dlssfr-dlsma}  and \ref{fig:dlssfr-dlsma-sdss}, this would be a strong trend and easily detected if present. Since no trend is seen, there must be another offsetting effect, but the only other knob in the model is to turn down the wind strength.  This could plausibly work in the right direction since more compact galaxies have higher $V_{esc}$ (at fixed \mstar) and would thus have weaker winds.

A quantitative estimate of the lifetime impact of concentration variations on star formation rates is available from previously unpublished data from the Dutton semi-analytic model \citep{Dutton-vandenBosch-Dekel-10}, which divides disk galaxies into annuli and calculates the local mass-loading factor $\eta$ at each radius.  Their winds are weaker in deeper potential wells and scale either as  $\eta \sim V_{esc}^{-1}$ (momentum-driven winds) or as $\eta \sim V_{esc}^{-2}$ (energy-driven winds). Data are available from the mocal for a collection of halos having energy-driven winds in a statistically realistic distribution of halo concentrations that are evolved appropriately over time. The scatter in \re\ at fixed \mstar\ is $\sim$0.25 dex, a good match to observations, but there is no trend in the results for smaller galaxies to have lower star formation rates, at either $z = 3$ or $z = 0$.  Smaller galaxies originate from higher-concentration halos, as expected, but their weaker winds fully cancel the effect of lower halo infall. This is a valuable simulation since it attempts to model the effects of wind and concentration over an entire galaxy's lifetime.  The lesson learned is that wind differences must be strong $--$ a parallel collection of galaxies with momentum-driven winds, which vary less with galaxy size, shows remaining correlated residuals in \delssfr\ vs.~\delsma.

This discussion brings us back to the EAGLE simulations, which exhibit many of the expected effects of halo concentration (and formation-time) differences. \citet{Matthee-17} study scatter about the stellar-mass-halo-mass relation, which they find to be strongly correlated with halo concentration: more concentrated halos form stars more rapidly, have weaker winds, and higher stellar mass at fixed halo mass \citep[see also][]{Wechsler-02, Zhao-09, Jeeson-Daniel-11, Ludlow-14, Correa-15,Kulier-19}. \citet{Matthee-17} note that initial concentration differences are amplified by the presence of baryons, which rapidly collect in the central region, further deepening the central potential.  Residuals about the SFMS relation are therefore even bigger when baryons are included.  \citet{Davies-19} and \citet{Oppenheimer-19} study the effects of concentration on the gas content of halos in EAGLE.  More tightly bound halos have higher concentration, earlier formation time, bigger black holes, lower gas content due to higher black hole feedback, and thus earlier quenching times.  Finally, \citet{Furlong-17} study residuals about the star formation main-sequence and find that smaller galaxies at fixed mass have \emph{lower} star formation rates today.  This agrees with the predicted lower accretion rates in high-concentration halos (cf.~Figure~\ref{fig:concentrationinfall}) but disagrees with our data showing no trend.  Perhaps the EAGLE wind prescription does not weaken winds enough in high-concentration halos.  We have not been able to find a discussion in the EAGLE literature explicitly treating the joint effects of concentration on galaxy radii and winds.

Finally we remind readers once again of the correlation between galaxy size and concentration in the NIHAO and VELA simulations \citep{Jiang-19}.  This motivated \citet{Chen-20} to posit concentration as the second halo parameter driving residuals in \reff\ vs.~\mstar.  Smaller (i.e., denser) galaxies make bigger black holes in their picture, which causes them to quench earlier in a manner similar to the EAGLE simulations.  The notion that halo concentration modulates black hole mass goes back to \citet{Booth-Schaye-10}, who noted that bigger black holes form in halos with higher binding energy in their simulations.  The cause in their case was weaker AGN feedback, not stellar feedback, but the same idea was present, namely, that halo concentration is a powerful second parameter influencing the life histories of galaxies.

\section{Conclusions}
\label{sec:conclusion}

In this work, we have investigated correlations between star formation rate and galaxy radius for star-forming galaxies $on$ the main-sequence. We have benefited from using large samples from the CANDELS and SDSS surveys, and our analysis covers a wide range of stellar mass and redshift. Since both SFR and \reff\ correlate with stellar mass and redshift, we remove these trends and study the residual correlations.  Our conclusions are as follows:

\begin{enumerate}

\item  In accordance with previous works, we confirm that there is no significant correlation between the star formation rate and \reff\ at fixed stellar mass for \mstar\ $\leq$ 10$^{10}$\msun.  This is true for both the CANDELS and SDSS samples at all redshifts.

\item A weak positive trend in star formation rate with \re\ appears above $10^{10}$\msun\ in CANDELS. The main cause seems to be the presence of small-radius galaxies well below the main-sequence, which are plausibly evolving slowly to the green valley.  These galaxies are visible in the SDSS sample also.

\item If \fgas\ were constant in all main-sequence galaxies at a given stellar mass, the Kennicutt-Schmidt and related density-dependent star formation laws would predict a strong upward trend in star formation rate towards smaller radius.  This trend is not seen, which means that smaller galaxies must have lower \fgas.

\item This prediction is confirmed by comparing to the measured gas contents of local galaxies in the xCOLD$\,$GASS and xGASS surveys.  The magnitude of the effect is about a factor of 2-3 from small to large galaxies.  

\item The lower gas fraction in smaller galaxies is consistent with the equilibrium bathtub model for galaxy evolution, in which the gas density adjusts itself to make stars at the rate mandated by the difference between the halo mass accretion rate and the mass loss rate due to winds.  In this reading, the star formation rate should be regarded as the independent variable in the KS law (set by halo minus wind), and the gas density is the dependent variable that results from applying the microphysics of the KS law.  Simply stated, small galaxies have less gas because their higher-density gas is more efficient at making stars.

\item Results from NIHAO, VELA, and EAGLE simulations suggest that halo concentration is an important second parameter in determining galaxy radius, \re, and that smaller galaxies form in higher-concentration halos.

\item Higher-concentration halos accrete more slowly at all masses \mstar\ $> 10^9$ \msun\ back to $z \sim 3$. Since small galaxies make stars at the same rate as large galaxies, this implies that stellar winds are weaker in deeper potential wells.

\end{enumerate}

We caution here that the bathtub model involves several simplifying assumptions, although previous studies shown that it can successfully predict many galaxy scaling relations \citep[e.g.,][]{Rodriguez-Puebla-16a}. For example, it does not describe the cycles between ISM and circumgalactic medium, which actually contribute a significant budget of baryons according to recent observations. Our prediction is a direct consequence of that assumption, and should be tested in more sophisticated simulations or future observations.

In summary, evidence is accumulating from several different directions that star-forming galaxies are a two-parameter family whose properties are set by halo mass and halo concentration.  The plot of \reff\ vs.~\mstar\ may be one of the clearest  mappings of this 2-D relationship.  At the same time, models suggest that halo concentration may modulate many other aspects of galaxy evolution as well, such as wind strength (and therefore composition), star formation rate, halo gas fraction, and black hole mass. All of these quantities are tightly interwoven throughout a galaxy's lifetime, and concentration is not perfectly constant over time. Stochastic short-term variations in halo mass accretion add further scatter, especially to star formation histories. An additional question is whether concentration is the right variable, or whether halo formation time is a better predictor. The complexity of the situation can therefore only be handled through simulations, but such simulations always have a number of free parameters, which are typically set by fitting to data.  We suggest that properly fitting the map of star formation rates in \re\ vs.~\mstar\ should be added to the standard arsenal of observations used to calibrate galaxy simulations.

\acknowledgments

We thank the referee for providing valuable comments, which significantly improved the manuscript. This work is based in part on observations taken by the CANDELS Multi-Cycle Treasury Program, which was supported under program number HST-GO-12060, provided by NASA through a grant from the Space Telescope Science Institute, which is operated by the Association of Universities for Research in Astronomy, Incorporated, under NASA contract NAS5-26555.  Members of the CANDELS team at UCSC acknowledge support from NASA \textit{HST} grant GO-12060.10-A and from NSF grants AST-0808133 and AST-1615730. LL acknowledges the support from the visiting program of Chinese Academy of Science and the National Science Foundation of China (No. U1831205, 11703063). AD was partly supported by NSF AST-1405962, BSF 2014-373 and GIF I-1341-303.7/2016. ZC acknowledges the NSFC grants  (11403016, 11433003, 11673018) and Innovation Program 2019-01-07-00-02-E00032 of SMEC. 

It is also based in part on data taken by the Sloan Digital Sky Survey. Funding for the SDSS and SDSS-II has been provided by the Alfred P. Sloan Foundation, the Participating Institutions, the National Science Foundation, the U.S. Department of Energy, the National Aeronautics and Space Administration, the Japanese Monbukagakusho, the Max Planck Society, and the Higher Education Funding Council for England. The SDSS Web Site is http://www.sdss.org/.

The SDSS is managed by the Astrophysical Research Consortium for the Participating Institutions. The Participating Institutions are the American Museum of Natural History, Astrophysical Institute Potsdam, University of Basel, University of Cambridge, Case Western Reserve University, University of Chicago, Drexel University, Fermilab, the Institute for Advanced Study, the Japan Participation Group, Johns Hopkins University, the Joint Institute for Nuclear Astrophysics, the Kavli Institute for Particle Astrophysics and Cosmology, the Korean Scientist Group, the Chinese Academy of Sciences (LAMOST), Los Alamos National Laboratory, the Max-Planck-Institute for Astronomy (MPIA), the Max-Planck-Institute for Astrophysics (MPA), New Mexico State University, Ohio State University, University of Pittsburgh, University of Portsmouth, Princeton University, the United States Naval Observatory, and the University of Washington.

The Rainbow database is operated by the Universidad Complutense de Madrid (UCM), partnered with the University of California Observatories at Santa Cruz (UCO/Lick,UCSC).

\bibliography{refs.bib}{}
\bibliographystyle{aasjournal}

\end{document}